# A Natural Cosmological Model with $\Omega_m <1$

George Chapline
Lawrence Livermore National Laboratory
Livermore, CA 94550

**Abstract**

Flat cosmological models with a cosmological constant on the order of the Einstein-de-Sitter critical mass density are enigmatic in the sense that there does not appear to be any natural explanation for why there should be a cosmological constant with such a magnitude. As a perhaps more natural alternative we propose a spatially homogeneous isotropic $\Omega=1$ universe containing cold matter and with a vacuum energy that varies inversely with the square of the Robertson-Walker scale parameter. This leads to a type of cosmological model that in some ways resembles yet is distinct from an open Friedmann universe with no cosmological constant. Some possibilities for distinguishing between a flat universe with a cosmological constant and a flat universe where the vacuum energy density is inversely proportional to the square of the scale parameter are discussed.

For many years now a dichotomy has existed between the bulk of astronomical and cosmochemical observations, which seemed to favor a low density open universe [Gott et. al. 1974], and aesthetic and theoretical prejudices, which have favored a flat ($\Omega=1$) universe [Kolb and Turner 1990]. This dichotomy has given rise to a minor industry : the search for "dark matter" on cosmological scales. On the other hand recent observations of Type I supernovae at cosmologically significant redshifts [Perlmutter et. al. 1998; Riess et. al. 1998] have tended to not only confirm the old suspicions that the average density of matter on cosmological distance scales is smaller than that required for a flat ($\Omega=1$) universe, but in addition these observations suggest that there is at the present time a positive vacuum energy whose magnitude is comparable to the average matter density. Since there are now strong hints from the way the amplitudes of primordial fluctuations in the cosmic microwave background vary with angular size that the universe is indeed flat [Lineweaver 1998], one can say that at the present time the problem of dark matter on cosmological scales has been resolved in favor of a vacuum energy density.

Unfortunately from the point of view of fundamental physics there does not appear to be any natural explanation for why there should be a vacuum energy density whose value at the present time just happens to be close to the Einstein-de Sitter critical density. In most field theories of elementary particles, for example, it is expected that there should be a vacuum energy density on the order of the 4th power of the Planck mass [Zel'dovich 1968]. It has been known of course for a long time that this is very many orders of magnitude larger than what would be allowed by observational cosmology. On the other hand it is amusing to note that the recently inferred value of the cosmological constant is on the order of the naive vacuum energy density divided by the square of the Robertson-Walker scale parameter expressed in Planck length units. Actually there are some compelling theoretical reasons for believing that in the fundamental theory underlying superstring theory not only is the universe locally conformally flat [Chapline 1992; 1993], but the vacuum energy density is non-zero and (in Planck units) on the order of the inverse square of the distance scale parameter for a Robertson-Walker metric [Chapline 1998]. This would imply that the cosmological "constant" is not really constant. Of course, the possibility of a time varying vacuum energy has always existed, and a priori there is no obvious reason why cosmological models with a vacuum energy varying inversely with the square of the scale parameter should be excluded.

The Raychaudhuri equation for a Robertson-Walker universe containing a cold matter density $\rho$ and where there is a nonzero vacuum energy density $\rho_{vac}$ reads

$$\frac{\ddot{R}}{R} = \frac{4\pi G}{3}(\rho + \rho_{vac} + 3p_{vac}) \ , \qquad (1)$$

where $p_{vac}$ is the vacuum pressure. We will assume that the vacuum energy density $\rho_{vac}$ satisfies an energy conservation equation of the usual form

$$\dot{\rho}_{vac} = -3(\rho_{vac} + p_{vac})\dot{R}/R \ , \qquad (2)$$

and that the cold matter density satisfies the usual conservation law for ordinary matter. When the energy densities present satisfy the usual energy conservation law, the Raychaudhuri equation can be integrated in the well known way [see e.g. Hawking and Ellis 1974] to yield an equation for the Hubble constant:

$$H^2 = \frac{8\pi}{3}G(\rho + \rho_{vac}) - \frac{\kappa c^2}{R^2}, \qquad (3)$$

where $H = \dot{R}/R$ is the Hubble constant at time t and $\kappa = \pm 1$ or $0$ specifies whether the universe is closed, open, or flat. Although the quantity $8\pi G\rho_{vac}$ looks like it plays exactly the same role as a cosmological constant $\Lambda$, it should be kept in mind that eq.(2) implies that when $\dot{\rho}_{vac} \neq 0$ it is no longer true that $p_{vac} = -\rho_{vac}$. Therefore when $\dot{\rho}_{vac} \neq 0$ it is not strictly speaking correct to identify the quantity $8\pi G\rho_{vac}$ as a "cosmological constant". For a flat universe where $8\pi G\rho_{vac}$ (cf. "$\Lambda$") $= \lambda/R^2$ we have

$$H^2 = \frac{8\pi}{3}G\rho + \frac{\lambda}{3R^2}. \qquad (4)$$

This equation resembles the equation for the Hubble constant in an open Friedmann universe, except that in an open Friedmann universe the Einstein field equations would require that $\lambda=1$. However eq. (4) can be integrated for any value of $\lambda$, yielding the following expressions for the Hubble constant and deceleration parameter q as a function of $R$ :

$$H = \frac{\sqrt{m}}{R^{3/2}}(1 + \frac{\lambda R}{3m})^{1/2} \qquad (5)$$

$$q = \frac{\Omega_m}{2}, \qquad (6)$$

where $m=(8\pi G\rho/3)R^3$ and $\Omega_m$ is the ratio of the density of ordinary matter to the critical density. At early times when $\Omega_\Lambda << \Omega_m$, H and q have values characteristic of an Einstein-de Sitter universe, while at late times when $\Omega_\Lambda >> \Omega_m$, H and q behave in a manner similar to that of an open Friedmann universe. The age of the universe in our cosmological model will be given by

$$t_0 = \frac{1}{H_0}[1 + \frac{1}{u^2} - \frac{\sqrt{1+u^2}}{u^3}\ln(u + \sqrt{1+u^2})], \qquad (7)$$

where $u^2 = \frac{\lambda R}{3m}$. The expressions (6) and (7) have exactly the same form as for an open Friedmann universe with no cosmological constant, except that for an open universe with no cosmological constant $u^2 = R/m$. At early times when $\Omega_\Lambda << \Omega_m$ $t_0$ is close to the

Einstein-de Sitter value $\frac{2}{3H_0}$, while at late times $t_0$ approaches the Milne value $H_0^{-1}$. Similarly at early times the deceleration parameter q is close to the Einstein-de Sitter value 0.5, while at late times q approaches zero. Thus the deceleration parameter and the relationship between the age of the universe and Hubble constant behave in a manner quite similar to that in an open Friedmann universe with no cosmological constant. Moreover, the predicted late time behavior of these quantities in our model with a time dependent vacuum energy density is quite different from what would be expected in any Robertson-Walker model with a time independent cosmological constant. For such models the deceleration parameter q approaches -1 and the Hubble constant becomes time independent.

If one could accurately measure the current value of q or $t_0$, it ought to be possible to distinguish whether we live in a flat universe with a time independent cosmological constant or as we have proposed one with a vacuum energy density varying as $1/R^2$. Unfortunately, it has not yet been possible to directly measure these quantities with sufficient accuracy to permit such a discrimination. At the present time the most reliable information regarding these cosmological parameters is provided by the recent measurements of the apparent brightness of distant supernovae. In our flat universe model with a vacuum energy varying as $R^{-2}$ the luminosity distance $d_L$ as a function of redshift will be given by

$$d_L = \frac{1}{H_0} (1+z) \, \Omega_\Lambda^{-1/2} \left[ \cosh^{-1}(2u_0^2 + 1) - \cosh^{-1}(\frac{2u_0^2}{z+1} + 1) \right] \qquad (8)$$

The observed abundance of deuterium suggests that at the present time $\Omega_m \approx 0.2$, which implies that at the present time $\Omega_\Lambda \approx 0.8$ and therefore $u_o^2 \approx 4$. For the most distant supernovae observed by Perlmutter, et. al. $z \approx 0.8$, in which case eq.8 using $\Omega_\Lambda \approx 0.8$ and $u_o^2 \approx 4$ predicts that $d_L = 1.18 \, H_0^{-1}$. By way of comparison in the Einstein-de Sitter universe one would have $d_L = 1.04 \, H_0^{-1}$. This prediction for the diminution of the apparent brightness of supernovae with $z \approx 0.8$ is consistent with the observations, but unfortunately is also close to what would be predicted by a model with a time independent cosmological constant [see e.g. Riess et. al. 1998].

Observations of supernovae at greater redshifts should eventually allow one to discriminate between a model with a constant $\Lambda$ and our model where the vacuum energy varies as $R^{-2}$. However a more immediate way to discriminate between these models may perhaps be provided by measurements of the probability of gravitational lensing of distant quasars by intervening galaxies. It is well known [Turner 1990] that the statistical probability of gravitational lensing of distant quasars is sensitive to the presence of a cosmological constant. In particular flat universe models with a time independent cosmological constant generally predict significantly larger probabilities for the lensing of large redshift quasars than models with no cosmological constant [Turner 1990]. Modeling the lensing galaxies as point masses yields the following formula for the lensing optical depth $\tau$ as a function of quasar redshift $z_Q$ [Turner et. al 1984]

$$\tau = \frac{3}{2} \Omega_L \int_1^{1+z} \frac{\varsigma_Q - \varsigma_L}{\varsigma_Q} \varsigma_L \frac{d(tH)}{dx} x^4 dx, \qquad (9)$$

where $\Omega_L$ is the cosmological density of lenses in units of the Einstein-de Sitter critical density, $x = 1 + z_L$ and $\varsigma_Q$, $\varsigma_L$ are proper time-like coordinates for the quasar and lensing object:

$$\varsigma = H_0 \int_{t_e}^{t_0} \frac{dt}{1+z}. \qquad (10)$$

In our cosmological model where the vacuum energy varies as $R^{-2}$ the affine parameter $\varsigma$ will be given by

$$\varsigma = (1 + u_0^2) \int_1^x \frac{dw}{w^3(w + u_0^2)}. \qquad (11)$$

To lowest order in $(1 + u_0^2)^{-1}$ the integral on the right hand side of (11) equals $1-x^{-2}$, in which case we obtain the following approximate formula for the lensing optical depth:

$$\tau = \frac{1}{2} \Omega_L \frac{z_Q^2}{2 + z_Q}. \qquad (12)$$

This formula is identical with the formula for the lensing optical depth due to point masses in a low density open universe obtained by Turner, Ostriker, and Gott [Turner et. al. 1984]. Although there is some uncertainty as to what would be a reasonable value for $\Omega_L$, the formula (12) appears to be in rough agreement with the observed frequency of lensed quasars. In contrast, the consistency of Robertson-Walker models with a time independent cosmological constant with these observations appears at the present time to be rather problematic [Falco et. al. 1998]. Fortunately our understanding of the nature of lensing galaxies is rapidly improving [see e.g. Kochanek et.al. 1998], and there is hope that this issue will be resolved in the near future.

The author would like to thank Michael Turner and Jim Peebles for help in understanding the current state of observational cosmology.

**References**

Chapline, G. 1992, A quantum model for space-time, Mod. Phys. Lett., A7, 1959.

Chapline, G. 1993, Anyons and coherent states for gravitons, *Proc. XXI International Conference on Differential Geometric Methods in Theoretical Physics*, ed. C.N. Yang, M. L. Ge and X. W. Zhou (Singapore: World Scientific).

Chapline, G. 1998, The vacuum enrgy in a condensate model for spacetime, hep-th/9812129.

Falco, E. E., Kochanek, C. S., and Munoz, J. A. 1998, Limits on cosmological models from radio-selected gravitational lenses, Ap. J., 494 , 47.

Gott J. R., Gunn, J. E., Schramm, D., and Tinsley B. M.1974, An unbound universe?, Ap. J., 194, 543.

Hawking, S. and Ellis, G. F. R. 1973, *The Large Scale Structure of Space-time* (Cambridge University Press).

Kochanek, C. S. et. al. 1998, Results from the Castles survey of gravitational lenses, astro-ph/9811111.

Kolb, E. R. and Turner, M. S. 1990,*The Early Universe*  (Addison-Wesley).

Lineweaver, C. H. 1998, The cosmic microwave background and observational convergence in the $\Omega_\Lambda$- $\Omega_m$ plane, Ap. J., 505, L69.

Perlmutter, S., et. al. 1998, Discovery of a supernovae explosion at half the age of the universe,  Nature, 391, 51.

Riess, A. G., et. al., 1998, Observational Evidence from supernovae for an accelerating universe and a cosmological constant, Ap. J., 116, 1009.

Turner, E. L., Ostriker, J. P., and Gott J. R. 1984, The statistics of gravitational lenses, Ap. J. 284, 1 .

Turner, E. L. 1990, ), Gravitational lensing limits on the cosmological constant in a flat universe, Ap J. 365, L43.

White S. D. M., et. al. 1993, The baryon content of galaxy clusters: a challenge to cosmological orthodoxy, Nature, 366, 429.

Zel'dovich, Ya. B. 1968, The cosmological constant and the theory of elementary particles, Usp. Fiz. Nauk, 95, 209.